\documentclass[journal]{IEEEtran}

\usepackage{graphicx}
\usepackage{multirow}
\usepackage{wrapfig}
\usepackage{subfigure}
\usepackage{amsmath}
\usepackage{cases}
\usepackage{array}
\usepackage{tikz}
\usepackage{tabularx}

\usepackage{algorithmic}
\usepackage{algorithm}
\usepackage{amsfonts}

\ifCLASSINFOpdf
\else
\fi

\begin{document}

\title{Acoustic Disturbance Detection for Autism Spectrum Disorder Diagnosis and Intelligibility Enhancement.}

\author{M. Pillonetto, A.~Queiroz,~\IEEEmembership{Student Member,~IEEE,}
        and~R.~Coelho,~\IEEEmembership{Senior~Member,~IEEE}\vspace{-0.4cm}% <-this % stops a space
        
\thanks{This work has been submitted to the IEEE for possible publication. Copyright may be transferred without notice, after which this version may no longer be accessible.}
\thanks{This work was supported in part by the National Council for Scientific and Technological Development (CNPq) 305488/2022-8 and Fundação de Amparo à Pesquisa do Estado do Rio de Janeiro (FAPERJ) under Grant 200518/2023 and in part by the Coordenação de Aperfeiçoamento de Pessoal de Nível Superior - Brasil (CAPES) - under Grant Code 001.}% <-this % stops a space
\thanks{The authors are with the Laboratory of Acoustic Signal Processing, Military Institute of Engineering (IME), Rio de Janeiro, RJ 22290-270, Brazil (e-mail: coelho@ime.eb.br).}% <-this % stops a space

}

\markboth{}%
{Shell \MakeLowercase{M. Pillonetto, A. Queiroz and R. Coelho}: Acoustic Disturbance Sensing Level Detection for Autism Spectrum Disorder Diagnosis and Intelligibility Enhancement}

% make the title  
\maketitle

\begin{abstract}

The acoustic sensitivity of Autism Spectrum Disorder (ASD) individuals highly impacts their intelligibility in noisy urban environments. In this paper, the disturbance sensing level is examined with perceptual listening tests that demonstrate the impact of their append High Internal Noise (HIN) profile on intelligibility. This particular sensing level is then proposed as additional aid to ASD diagnosis. In this paper, a novel intelligibility enhancement scheme (ISE$_\text{ASD}$ - Intelligibility Sensing Enhancement for ASD individuals) is also introduced for ASD particular circumstances. For this proposal, harmonic features estimated from speech signal with HHT-Amp (Amplitude-based Hilbert Huang Transform) are considered as center frequencies of auditory filterbanks. A gain factor is further applied to the output of the filtered samples to emphasize the harmonic components of speech reducing the masking effects of acoustic noises. The proposed solution is compared to three competing approaches considering four acoustic noises at different signal-to-noise ratios. Two objective measures (ESTOI and PESQ) are also adopted for evaluation. The experimental results show that the personalized solution outperformed the competing approaches in terms of intelligibility and quality improvement.

\end{abstract}


\IEEEpeerreviewmaketitle

\section{Introduction}

\IEEEPARstart{A}{utism} Spectrum Disorder (ASD) is a neurodevelopmental disorder that affects approximately one in every hundred people around the world \cite{LEE_2024}\cite{GAO_2024}. It is denoted by difficulties in social interaction and the presence of repetitive behaviors \cite{LALAWAT_2024}. Auditory hypersensitivity is a key factor that characterizes autism \cite{RAJAN_2016}\cite{PARK_2017} and impacts speech communication in ASD individuals. This disturbance sensing is even more challenging in urban environments with different acoustic noisy conditions. This is a crucial cause of impairment and a key challenge for the acoustic signal processing research area. These effects may cause distortions on speech components e.g., the formants of the fundamental frequency (F0) \cite{FALK_2013}\cite{BONE_2017}\cite{QUEIROZ_2022}\cite{SHUKLA_2023}, and signal harmonic region related to the acoustic intelligibility \cite{COOKE_2009}\cite{TAO_2010}. This concern underlies several applications and systems such as robot audition, hearing aids, speech and speaker recognition.

Research works have been proposed to investigate and characterize autism behaviors. In \cite{LEE_2024}, conversation-level and acoustic-prosodic features were taken into account in a ASD classification task. Particularly, Vowel Space Characteristics (VSC) measurements improved the accuracy results in both ASD classification and severity score regression \cite{LEE_2024}. The relationship between these autistic traits and speech perception are also examined in \cite{IMAIZUMI_2024}. Studies using noisy speech tests have demonstrated that autistic individuals have lower ability of speech perception than Neurotypical (NT) \cite{ALCANTARA_2004}\cite{SCHELINSKI_2020}. Thus, background noise is a crucial issue that impacts communication of people with ASD.

Several signal processing methods are described in the literature to deal with noise interference. Speech enhancement approaches \cite{HENDRIKS_2012} \cite{FLANDRIN_2014} estimate the noise characteristics in order to attenuate such effects for quality assessment. However, this achievement not necessarily lead to speech intelligibility improvement \cite{KIM_2011}. On the other hand, acoustic masks \cite{LI_2009}\cite{LOIZOU_2010}\cite{FARIAS_2021} are defined to emulate the \textit{cocktail party} effect. These solutions provide intelligibility enhancement for the target speech signal.

Recently, the investigation of harmonic components from noisy speech signals has gained significant attraction \cite{EALEY_2001}\cite{WANG_2017}. Particularly, in strategies developed to achieve intelligibility gain \cite{QUEIROZ_2024}. In \cite{GAEL_2016} the formant center frequencies from voiced speech segments are shifted away from the region of noise \cite{GAEL_2017}, emulating the Lombard effect \cite{LOMBARD_1911}. Results showed that the Smoothed Shifting of Formants for Voiced segments (SSFV) is able to improve the intelligibility of speech signals in car noise environment. A different approach in \cite{NORHOLM_2016} used the amplitude and phase estimation filter \cite{STOICA_1999} with the harmonic models (APES$_\text{HARM}$ - Harmonic-based Amplitude and Phase EStimation). This procedure led to improved signal-to-noise ratio (SNR) of the reconstructed speech signals. Finally, the F0-based Gammatone Filtering (GTF$_\text{F0}$) method \cite{QUEIROZ_2021} is described to attain intelligibility enhancement. This scheme uses F0 estimates \cite{HHT} as center frequencies of auditory filterbank. Then, output samples are amplified by a gain factor attenuating the noise masking effect, which achieves intelligibility improvement.

However, in the specific case of people with ASD a key issue must be addressed: the presence of intense internal noise (High Internal Noise - HIN) \cite{PARK_2017}. The HIN is a crucial impairment for the daily social communication of individuals with ASD \cite{SHUKLA_2023}. It is characterized by the amplification of noise interference in the inner ear. Specifically, the elevated internal noise in ASD may produce atypically large fluctuations in neural responses, leading to unreliable and less predictable representations of the environment \cite{PARK_2017}. The composition of HIN with the external noise affects the F0 and formant of the speech signal in the ASD context \cite{BONE_2017}. This effect can also be perceived in the articulatory acoustic trait of natural spoken interactions of individuals with ASD symptoms \cite{LEE_2024}. Thus, a strategy to improve intelligibility must mitigate the effects of external noise by highlighting the harmonic frequencies of speech signal.

This work proposes a Gammatone Filtering ISE$_\text{ASD}$ (Intelligibility Sensing Enhancement for ASD individuals) method especially to attain intelligibility enhancement for ASD individuals. This solution is composed of three steps: initially, HHT-Amp is adopted to estimate the fundamental frequency from the noisy speech signal. The F0 information and its harmonic are considered as center frequencies of a Gammatone filterbank in the second stage. Finally, the filtered components are amplified by a gain factor to highlight the harmonic components of speech. This amplification mitigates the masking effects of background noise. This strategy aims to increase the robustness of external noise filtering for ASD condition leading to intelligibility improvement. This work also includes perceptual tests with neurotypical and ASD volunteers as an auxiliary instrument to detect Autism Spectrum Disorder.

Extensive experiments are conducted to evaluate the effectiveness of the ISE$_\text{ASD}$ for speech intelligibility and quality improvement. For this purpose, speech utterances collected from the TIMIT \cite{TIMIT_1993} database are corrupted by four real acoustic noises, considering four SNR values: -10 dB, -5 dB, 0 dB, and 5 dB. The proposed method and three baseline approaches are examined in terms of intelligibility and quality assessment. To this end, the Extended Short-Time Objective Intelligibility (ESTOI) \cite{JENSEN_2016} and Perceptual Evaluation of Speech Quality (PESQ) \cite{PESQ_2001} objective measures are adopted. The objective results show that the proposed solution outperforms the competitive methods in terms of speech intelligibility and quality. Perceptual tests are conducted with NT and ASD volunteers to verify the subjective intelligibility gain attained by the proposed solution.

The main contributions of this study are as follows:

\begin{itemize}
 \item Introduction of the ISE$_\text{ASD}$ method to improve the intelligibility considering the revealed HIN disorder of the ASD individuals;
 \item Definition of an auxiliary strategy for ASD diagnosis based on perceptual tests;
 \item Interesting speech intelligibility and quality improvement even for challenging noisy conditions;
 \item Additional perceptual tests to demonstrate that ISE$_\text{ASD}$ outperforms the baseline methods.

\end{itemize}

The remainder of this paper is organized as follows. An explanation of the High Internal Noise profile in ASD individuals is provided in Section II. Section III describes the steps of the proposed ISE$_\text{ASD}$ method for intelligibility enhancement. Section IV presents the competitive methods SSFV, APES$_\text{HARM}$ and GTF$_\text{F0}$, followed by the evaluation experiments and results. Finally, Section V concludes this work.

\section{Intelligibility for Autism Spectrum Disorder Situation}

\begin{figure}[t!]
 \centering
 \includegraphics[width=0.93\linewidth,clip=true,trim=0pt 0pt 0pt 0pt]{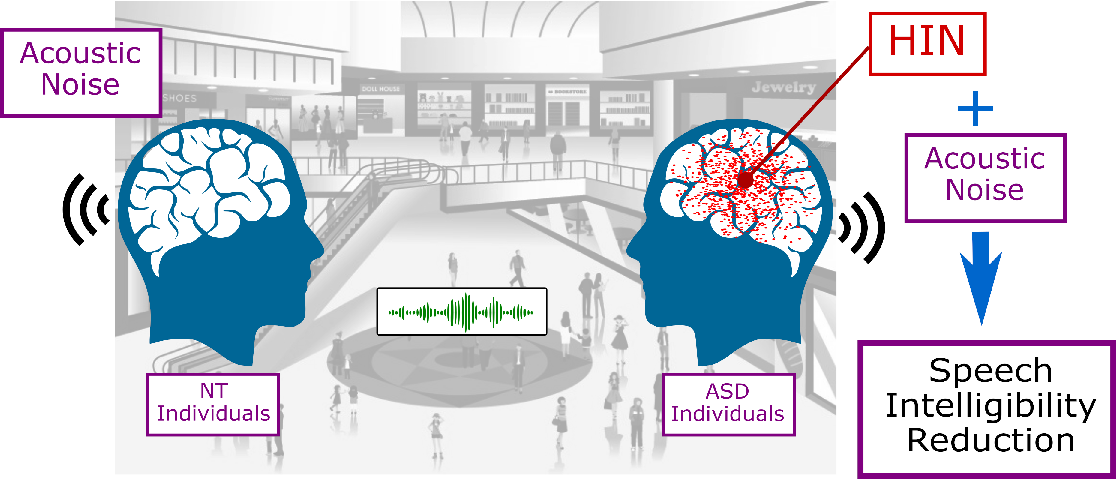}
 \caption{(a) Illustration of the High Internal Noise impact in the acoustic intelligibility of ASD individuals.}
 \label{asdEff}
\end{figure}

\begin{figure}[t]
 \centering
 \includegraphics[width=0.85\linewidth,clip=true,trim=0pt 0pt 0pt 0pt]{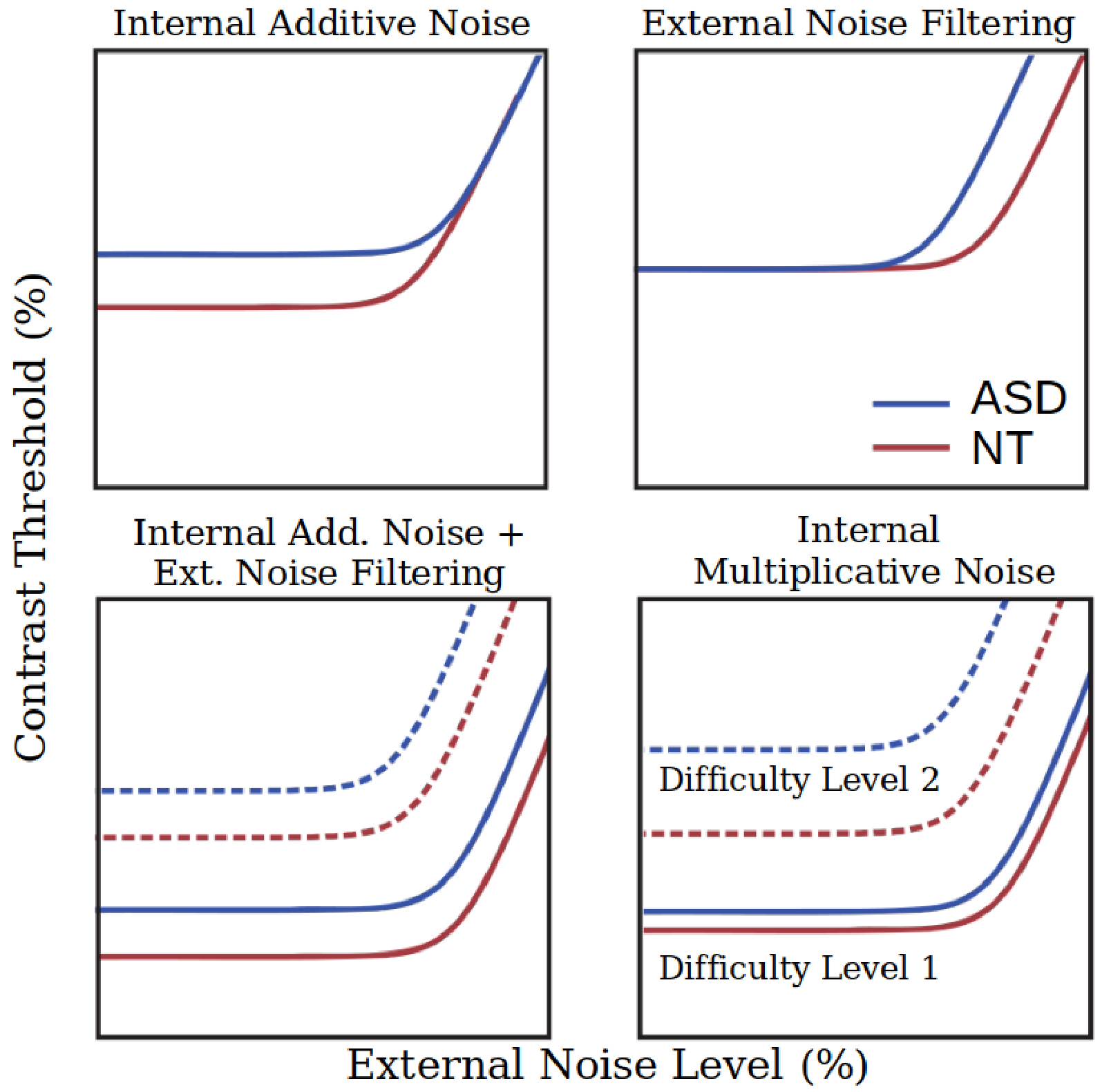}
 \caption{Perceptual Threshold versus Noise (TvN) for ASD versus NT individuals \cite{PARK_2017}.}
 \label{internal_noise}
\end{figure}

Autism is characterized by deficit in the behavioral/social and sensory fields\footnote{Resolution on Autism Specter Disorder (WHA 67.8)} and acoustic perceptual skills \cite{SHUKLA_2023}. This condition is typical in the syndrome, and audiometric tests are already adopted as an auxiliary instrument for the disorder detection. It is important to highlight that population with ASD are considered normal-hearing, i.e., it is assumed that individuals are not affected by hearing impairment \cite{RAJAN_2016}. An emerging hypothesis considers that a High Internal Noise (HIN) profile \cite{PARK_2017} plays an important role for the degradation of acoustic perception in ASD. Fig. \ref{asdEff} depicts the HIN effect in an acoustic noisy environment, whose equivalent degradation can be quantified by the sum of internal and external interference.

\begin{figure*}[t!]
\centering
\includegraphics[width=.93\linewidth,clip=true,trim=0pt 0pt 0pt 0pt]{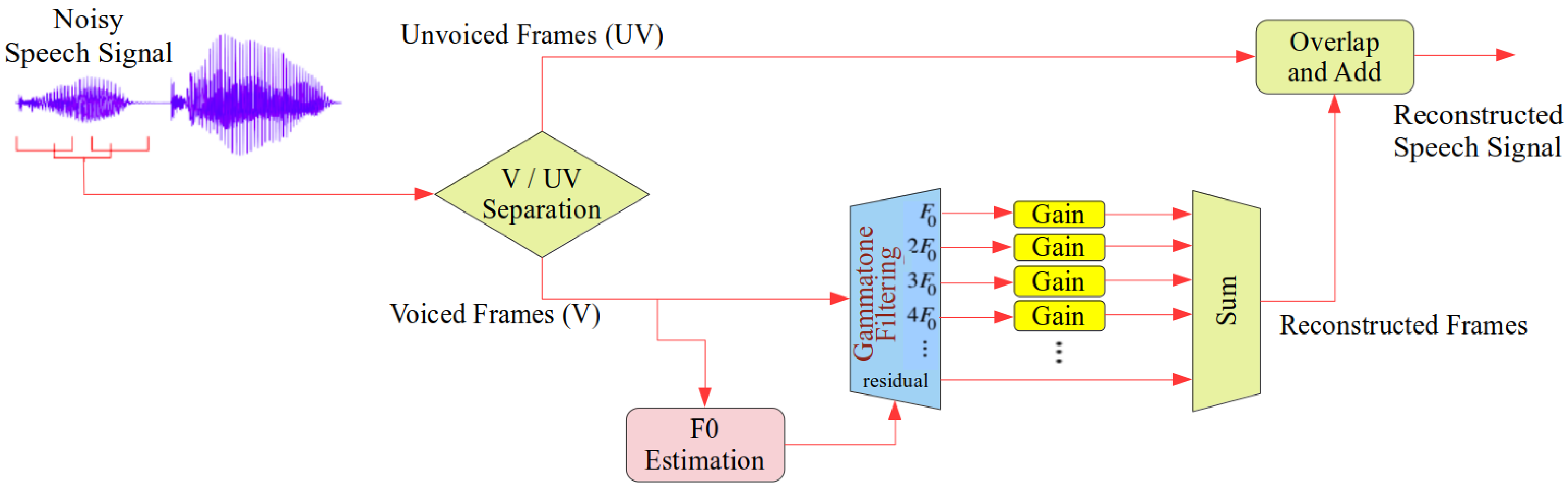}
\caption{Block diagram of the proposed ISE$_\text{ASD}$ method for speech intelligibility gain for ASD individuals.}
\label{schematic}
\end{figure*}

Experimental results \cite{PARK_2017} indicate the HIN profile with filtering of external noisy signals. The equivalent degradation can be quantified by the sum of internal and external interference. Perceptual sensitivity is measured in \cite{PARK_2017} at various levels of additive external noise. These levels are presented as measures of TvN (Threshold versus Noise), which allow the effects of noise to be quantified in NT and ASD individuals. Fig. \ref{internal_noise} shows TvN functions \cite{PARK_2017}, whose curves in these graphs follow a pattern: a segment parallel to the horizontal axis followed by an ascending segment. The contrast threshold is represented on the vertical axis, while the level of external noise is represented on the horizontal axis. In the upper left subplot of Fig. \ref{internal_noise}, the additive internal noise without filtering of autistic is compared with neurotypicals. Observe that ASD people have a greater presence of naturally additive internal noise when compared to the rest of the population. 

In the graph on the top right, it can be seen the TvN of the external noise filtering for the NT and ASD cases. This feature in combination with HIN should yield increased thresholds across all levels of external noise (bottom left). A similar pattern is expected if internal multiplicative noise is elevated for different difficulty levels (bottom right). Either HIN and external filtering may cause more negative end-of-sentence pitch slopes and less typical harmonics-to-noise ratio \cite{BONE_2017}. Therefore, individuals with ASD have lower perceptual intelligibility at loud external noise when compared to neurotypical in the same conditions.

% In \cite{HHT}, it was shown that the HHT-Amp method achieves interesting results in estimating the fundamental frequency of noisy speech signals. The HHT-Amp was evaluated in a wide range of noisy scenarios, including five acoustic noises with different nonstationarity degrees. It outperformed four competing estimators in terms of gross error (GE) and mean absolute error (MAE).

\section{The proposed ISE$_\text{ASD}$ Method}
\label{sec:proposal}

The ISE$_\text{ASD}$ introduces a set of gain factors to the harmonic \cite{QUEIROZ_2021} components to attenuate the HIN in the particular case of ASD individuals \cite{PARK_2017}. Fig. \ref{schematic} illustrates the block diagram of the proposed scheme. It can be separated in three main steps: F0 estimation, Gammatone filtering and samples amplification with gain factor. The target noisy signal $x(t)$ is first split into $Q$ overlapping 32 ms voiced/unvoiced (V/UV) frames $x_q(t), q = 1, 2, \ldots, Q$, with $50\%$ overlapping. In this work V/UV separation was previously applied two define two disjoint sets: $S_v$ is formed by frames that contain voiced speech, and $S_u$ is composed of the remaining segments, i.e., unvoiced speech and noise only.

\subsection{F0 Estimation: HHT-Amp}

The HHT-Amp method \cite{HHT} applies the Hilbert-Huang transform (HHT) \cite{HUANG_98}  to analyze the target speech signal. Different from other HHT-based approaches, the F0 value is not estimated from the instantaneous frequencies of the target signal. Instead, the fundamental frequency is obtained from the instantaneous amplitude functions. The HHT-Amp method is summarized as follows:
\begin{enumerate}
\item Apply the Ensemble Empirical Mode Decomposition (EEMD) \cite{HUANG_98} to decompose the voiced sample sequence $x_q(t)$ into a series of intrinsic mode functions (IMF) and a residual $r_q(t)$, such as:

\begin{equation}
 x_q(t) = \sum_{k=1}^K \mbox{IMF}_{k,q}(t) + r_q(t).
\end{equation}

\item Compute the instantaneous amplitude functions

\begin{equation}
 a_{k,q}(t) = |Z_{k,q}(t)|, k = 1, \ldots, K,
\end{equation}
from the analytic signals defined as

\begin{equation}
 Z_{k,q}(t) = \mbox{IMF}_{k,q}(t) + j \, H\{\mbox{IMF}_{k,q}(t)\},
\end{equation}
where $H\{\mbox{IMF}_{k,q}(t)\}$ refers to the Hilbert transform of $\mbox{IMF}_{k,q}(t)$.

\item Calculate the ACF 

\begin{equation}
 r_{k,q}(\tau) = \sum_t  a_k(t) \, a_k(t+\tau)
\end{equation}
of the amplitude functions $a_{k,q}(t), k = 1, \ldots, K$.

\item For each decomposition mode $k$, let $\tau_0$ be the lowest $\tau$ value that correspond to an ACF peak, subject to $\tau_{min} \leq \tau_0 \leq \tau_{max}$. The restriction is applied according to the range $[F_{min}, F_{max}] = [50,400] Hz$ of possible $F0$ values. The $k$-th pitch candidate is defined as $\tau_0 / {f_s}$, where $f_s$ refers to the sampling rate.

\item Apply the decision criterion defined in \cite{HHT} to select the best pitch candidate $\hat T_0$. The estimated $F0$ is given by $\hat F0 = 1 / \hat T_0$.

\end{enumerate}

In \cite{HHT}\cite{QUEIROZ_2022}, it was shown that the HHT-Amp method achieves interesting results in estimating the fundamental frequency of noisy speech signals. The HHT-Amp was evaluated in a wide range of noisy scenarios, including five acoustic noises with different nonstationarity degrees. It outperformed four competing estimators in terms of Gross Error (GE) and Mean Absolute Error (MAE).

\begin{figure*}[t!]
\centering
\includegraphics[width=0.33\linewidth,clip=false,trim=0pt 0pt 0pt 0pt]{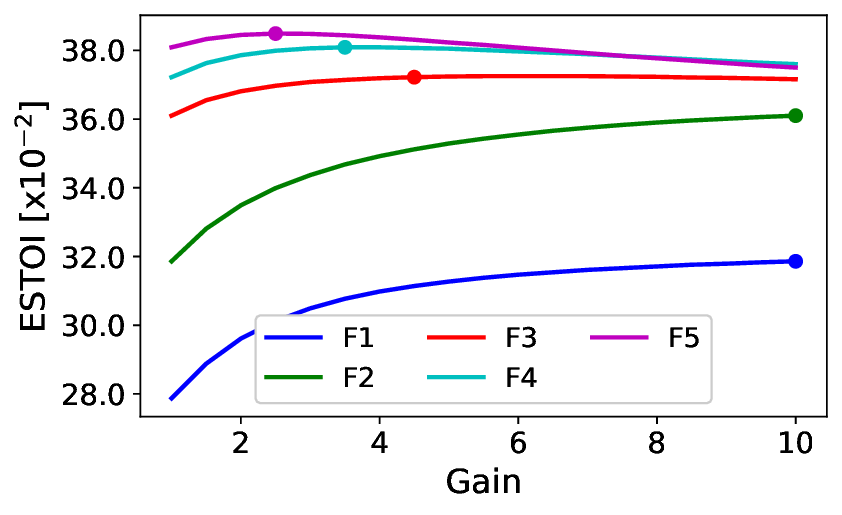}\hspace{-0.2cm}
\includegraphics[width=0.33\linewidth,clip=false,trim=0pt 0pt 0pt 0pt]{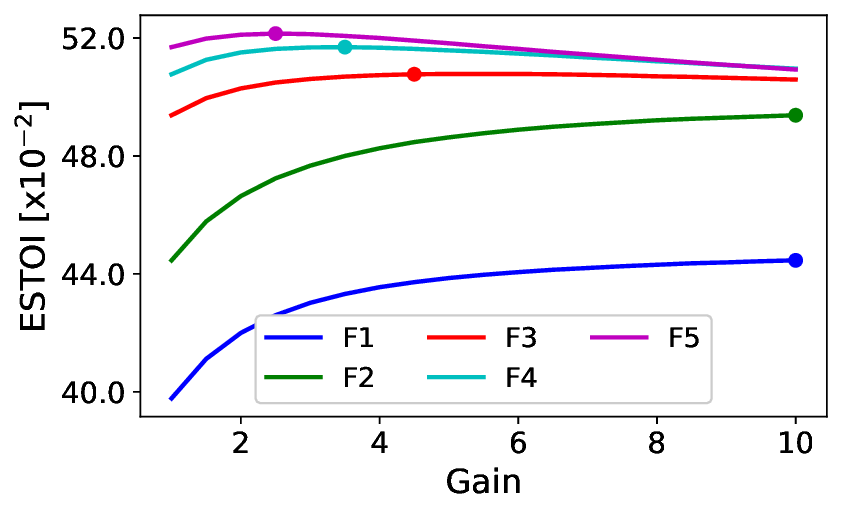}\hspace{-0.2cm}
\includegraphics[width=0.33\linewidth,clip=false,trim=0pt 0pt 0pt 0pt]{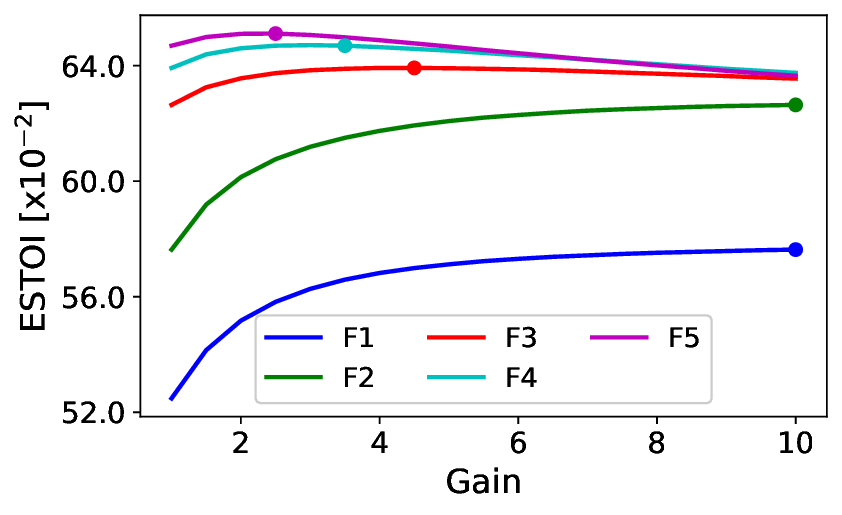}

\subfigure[]{\includegraphics[width=0.33\linewidth,clip=false,trim=0pt 0pt 0pt 0pt]{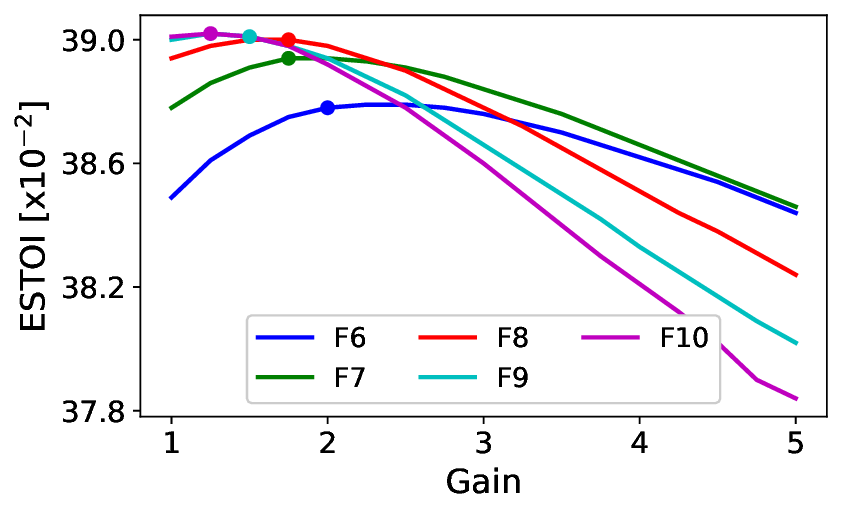}}\hspace{-0.2cm}
\subfigure[]{\includegraphics[width=0.33\linewidth,clip=false,trim=0pt 0pt 0pt 0pt]{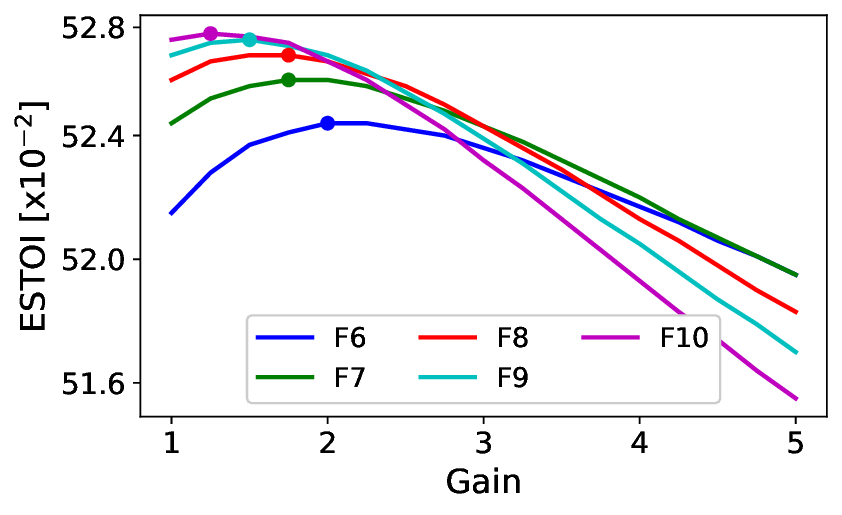}}\hspace{-0.2cm}
\subfigure[]{\includegraphics[width=0.33\linewidth,clip=false,trim=0pt 0pt 0pt 0pt]{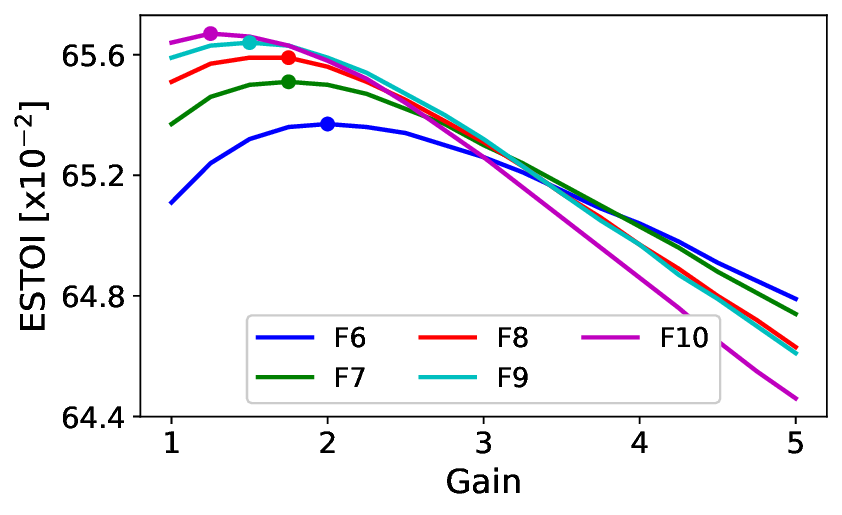}}
\caption{ESTOI average scores to characterize the personalized gains for speech signals corrupted by SSN noise with SNR values: (a) -5 dB, (b) 0 dB and (c) 5 dB.}
\label{gain_curves}
\end{figure*}

\subsection{Gammatone Filtering}

For each voiced frame $q \in S_v$, a set of Gammatone filters are used to filter the sample sequence $x_q(t)$. The time-domain impulse response of the Gammatone filter \cite{JOHANNESMA_1972} is defined as

\begin{equation}
g(t) = a t^{n-1} \cos(2 \pi f_c t + \phi) e^{-2 \pi b t}\, , \, t \geq 0\, ,
\label{eq:gamma}
\end{equation}
where $a$ is the amplitude, $n$ is the filter order, $f_c$ is the center frequency, $\phi$ is the phase, and $b$ is the bandwidth. In \cite{PATTERSON_1992}, it was shown that a set of fourth-order Gammatone filters are able to represent the magnitude characteristic of the human auditory system.

In the proposed method, a set of $L$ Gammatone filters $\{h_k(t),k=,\cdots,L\}$ are implemented \cite{COOKE_1993} considering order $n = 4$ and bandwidth $b = 0.25 \hat{F}0$. The center frequencies are set to 

\begin{equation}
 f_c = \hat{F}0, 2 \hat{F}0, \ldots, L \hat{F}0
\end{equation}
where $\hat{F}0$ is the pitch value estimated by the HHT-Amp method \cite{HHT}.
In order to align the impulse response functions, phase compensation is applied to all filters, which correspond
to the non-causal filters

\begin{equation}
h_k(t) = a (t+t_c)^{n-1} \cos(2 \pi f_c t) e^{-2 \pi b (t+t_c)}\, , \, t \geq -t_c\, ,
\label{eq:gamma_c}
\end{equation}
where $t_c = \frac{n-1}{2 \pi b}$, which ensures that peaks of all filters occur at $t = 0$.

Let $x_q^0(t) = x_q(t)$, the filtered signals $y_q^k(t), k = 1, \ldots, L$, are recursively computed by

\begin{equation}
\left\{\begin{array}{l}
y_q^k(t) = x_q^{k-1}(t) * h_k(t) \vspace{0.2cm}\\
x_q^k(t) = x_q^{k-1}(t) - y_q^k(t)
\end{array} \right. ,
\quad k = 1, \ldots, L\, .
\label{eq:filters}
\end{equation}

The residual signal defined as $r_q(t) = x_q^L(t)$ to guarantee the completeness of the input sequence.

\subsection{Gain Factors Characterization}

After the Gammatone filtering, the amplitude of the filtered samples $y_q^k(t), k = 1, \ldots, L$, are amplified by a gain factor $G_k \geq 1$. The main idea is to emphasize the presence of the integer multiple harmonics of the fundamental frequency, which will lead to speech intelligibility improvement, without introducing any noticeable distortion to the speech signal. The reconstruction of the voiced frame $q \in S_v$ leads to the sample sequence

\begin{equation}
\hat x_q(t) =  \sum_{k=1}^L G_k \, y_q^k(t) + r_q(t)\, .
\label{eq:frame_rec}
\end{equation}

The $G_k$ is defined as the values that better improves the ESTOI \cite{JENSEN_2016} intelligibility rate $d$ of a set of test signals. It is initially set as $\hat{G}_k = 1$, whose value is increased by $\Delta \hat{G} = 0.25$. If any gain lead to an increase in the average ESTOI ($D$), $G_k$ is selected according to the maximum intelligibility score, i.e.,

\begin{equation}
G_k \leftarrow \hat{G}_k = \mathrm{arg} \underset{1 \leq G \leq 10}{\mathrm{max}} D.
\end{equation}

In this work, the gains are characterized considering a training subset of 72 speech signals of the TIMIT database defined in \cite{GONZALEZ_2014}. The speech utterances are corrupted by Speech Shaped Noise (SSN) from DEMAND \cite{DEMAND_2013} database with three SNR values: -5dB, 0dB and 5dB.

Fig. \ref{gain_curves} illustrates the ESTOI objective intelligibility scores for noisy speech signals. The gain configuration starts from the first filter (F1), and its value is incremented until ESTOI reaches its maximum value (highlighted point). This gain is fixed, and the process is repeated for the subsequent filter. Observe that the same set of gains is attained for the three SNR values. After the process, the personalized gains that lead to the highest intelligibility ESTOI scores were defined for the first 10 filters. The achieved values of ISE$_\text{ASD}$ are defined as

\begin{equation}\label{gain_sets}
 G_k = \{ 10, 10, 4.5, 3.5, 2.5, 2, 1.75, 1.75, 1.5, 1.25 \}.
\end{equation} 
This set turns the harmonic components of speech more prominent when compared to the noisy signal. This effect may reduce the impact of the acoustic noise, and consequently, improve speech intelligibility.

The final step of the ISE$_\text{ASD}$ method is the reconstruction of the entire speech signal. For this purpose, the reconstructed voiced frames in $S_v$ and all the remaining frames in $S_u$ are joined together keeping the original frames indices. Thus, all frames are overlap and added to reconstruct the modified version $\hat x(t)$ of the target speech signal. The proposed ISE$_\text{ASD}$ method is summarized in Algorithm \ref{alg:alg1}.

\begin{algorithm}[t!]
\caption{ISE$_\text{ASD}$ Scheme.}\label{alg:alg1}
\begin{algorithmic}
\STATE\textbf{for} $q$ \textbf{do}
\STATE\setlength{\leftskip}{0.5cm}Input: $x_q(t)$
\STATE\textbf{F0 Estimation}
\STATE$\hat{F}0_q \gets$ with HHT-Amp \cite{HHT} as in Section III-A.
\STATE\textbf{Gammatone Filtering}
\STATE\textbf{for} $k$ \textbf{do}
\STATE\setlength{\leftskip}{1cm}$h_k(t) \gets$ impulse response of non-causal filters (\ref{eq:gamma_c})
\STATE$y_q^k(t) = x_q^{k-1}(t) * h_k(t)$
\STATE$x_q^k(t) = x_q^{k-1}(t) - y_q^k(t)$
\STATE\setlength{\leftskip}{0.5cm}\textbf{end for}
\STATE$y_q^k(t) * G_k \gets$ emphasize harmonics with $G_k$ as in (\ref{gain_sets})
\STATE$r_q(t) = x_q^L(t) \gets$ residual components
\STATE$\hat x_q(t) \gets$ voiced frames reconstruction as in (\ref{eq:frame_rec}).
\STATE$\hat x(t) \gets$ overlap and add technique.
\STATE\setlength{\leftskip}{0cm}\textbf{end for}
\STATE\textbf{return} $\hat x(t)$
\end{algorithmic}
\label{alg1}
\vspace{0.2cm}
\end{algorithm}

% \begin{figure*}[t!]
% \centering
% \includegraphics[width=0.45\linewidth,clip=false,trim=0pt 0pt 0pt 0pt]{figs/leg_destoi.eps}\\\vspace{0.2cm}
% \includegraphics[width=0.50\linewidth,clip=false,trim=0pt 0pt 0pt 0pt]{figs/destoi_snr-10.eps}\hspace{-0.2cm}
% \includegraphics[width=0.50\linewidth,clip=false,trim=0pt 0pt 0pt 0pt]{figs/destoi_snr-5.eps}\vspace{0.4cm}
% \includegraphics[width=0.50\linewidth,clip=false,trim=0pt 0pt 0pt 0pt]{figs/destoi_snr0.eps}\hspace{-0.2cm}
% \includegraphics[width=0.50\linewidth,clip=false,trim=0pt 0pt 0pt 0pt]{figs/destoi_snr5.eps}
% \caption{ESTOI improvement [$\times10^2$] boxplots for the noisy speech signals in different SNR values processed with the competitive methods: SSFV, APES$_\text{HARM}$, GTF$_\text{F0}$ and ISE$_\text{ASD}$.}
% \label{estoi_boxplots}
% \end{figure*}

\begin{figure*}[t!]
\centering
\includegraphics[width=0.45\linewidth,clip=false,trim=0pt 0pt 0pt 0pt]{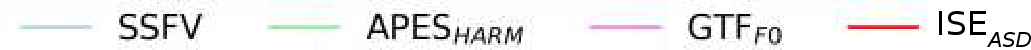}\\\vspace{0.2cm}
\includegraphics[width=0.50\linewidth,clip=false,trim=0pt 0pt 0pt 0pt]{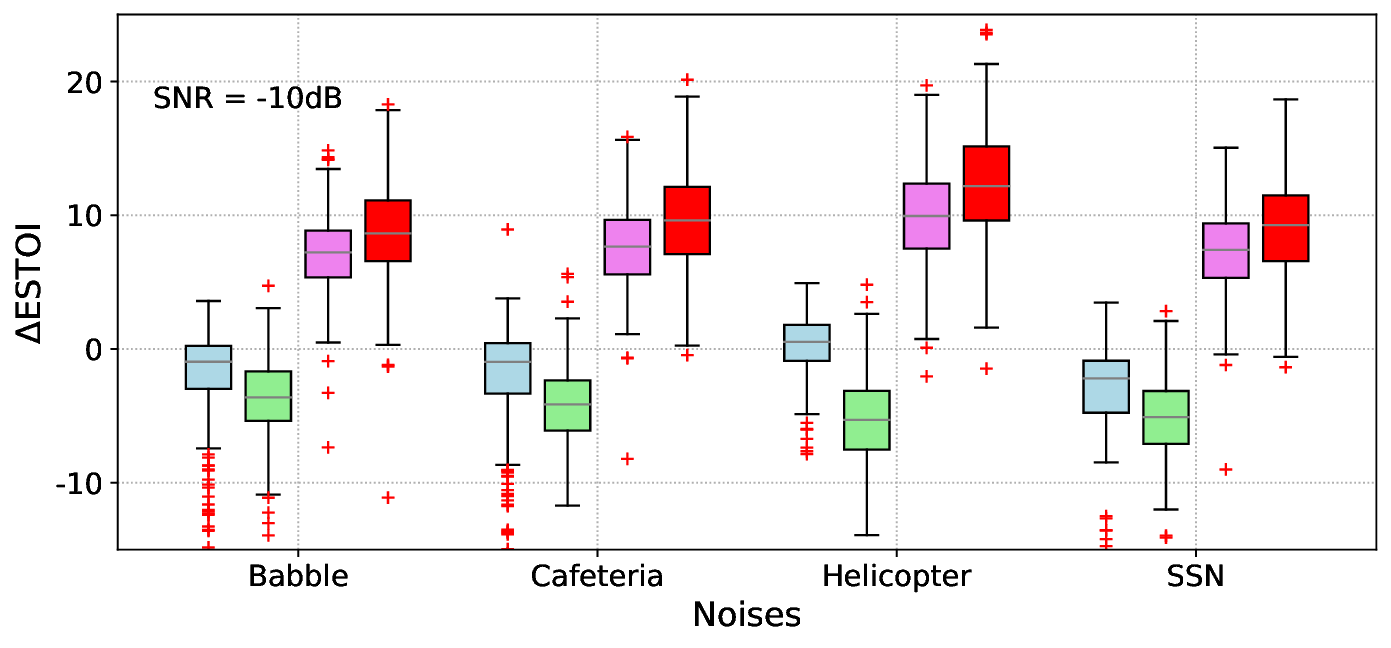}\hspace{-0.2cm}
\includegraphics[width=0.50\linewidth,clip=false,trim=0pt 0pt 0pt 0pt]{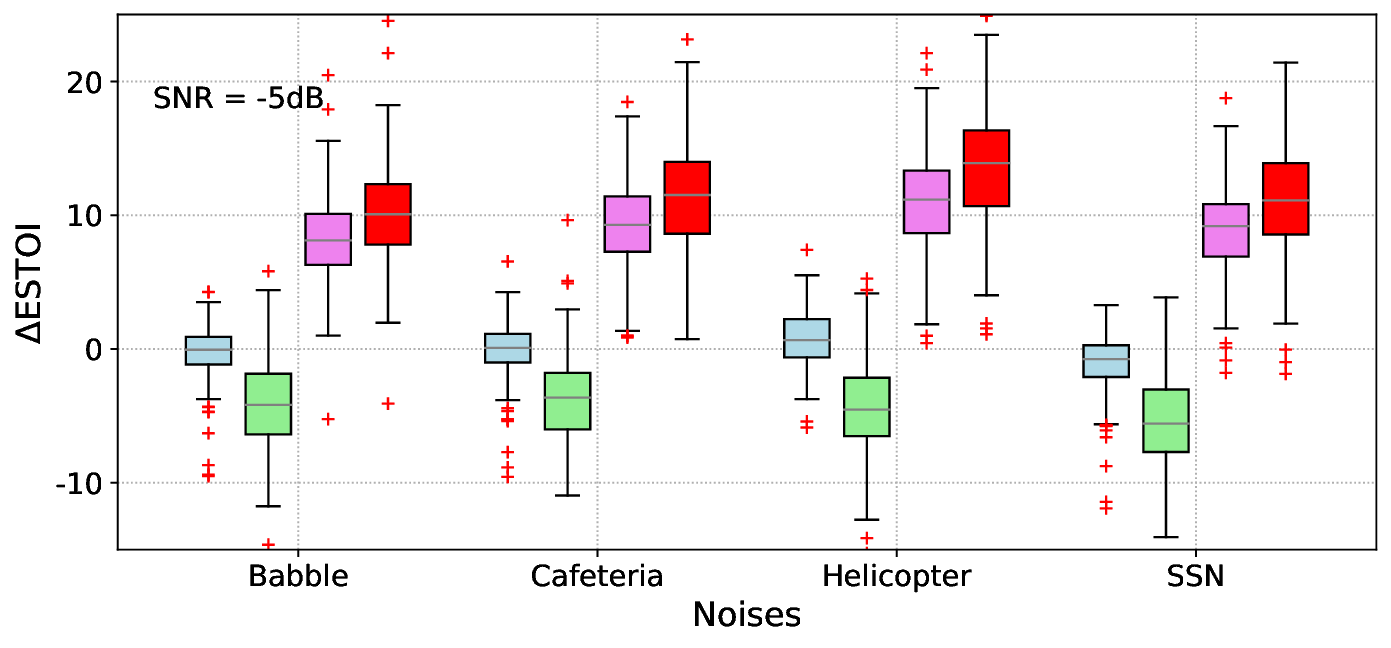}\vspace{0.3cm}
\includegraphics[width=0.50\linewidth,clip=false,trim=0pt 0pt 0pt 0pt]{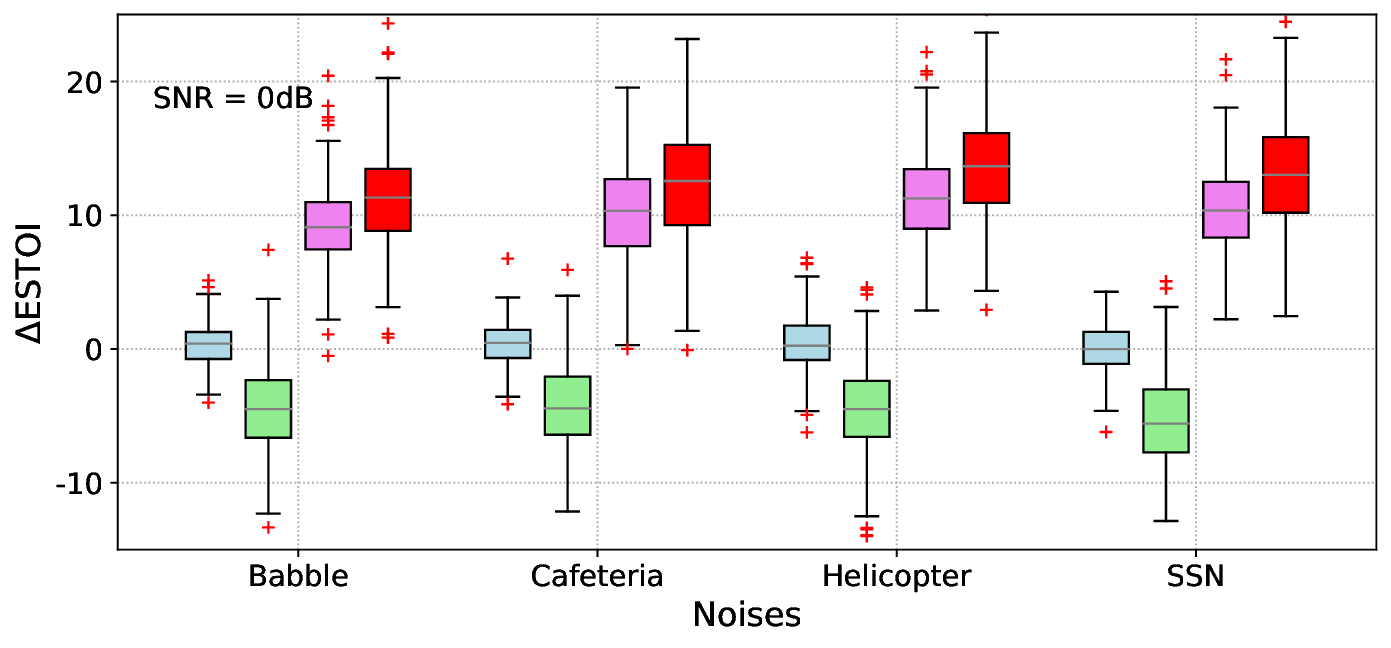}\hspace{-0.2cm}
\includegraphics[width=0.50\linewidth,clip=false,trim=0pt 0pt 0pt 0pt]{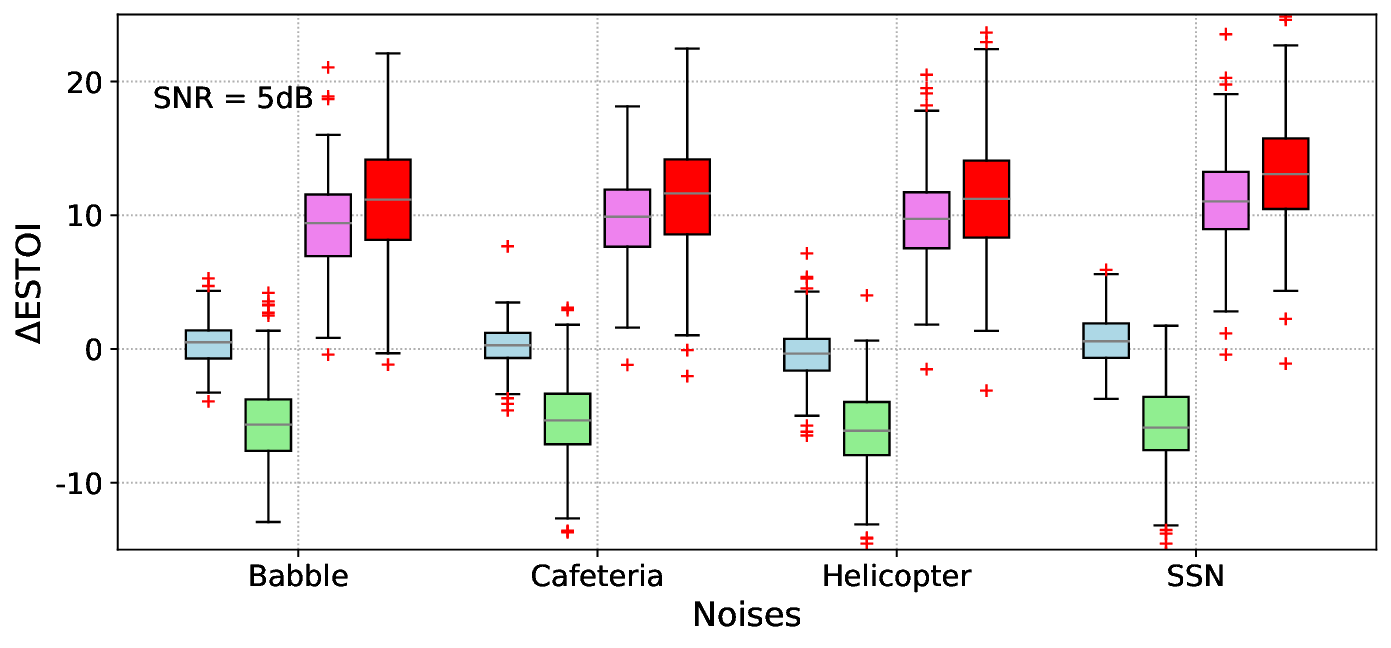}
\caption{ESTOI improvement [$\times10^{-2}$] boxplots for the noisy speech signals in different SNR values processed with the competitive methods: SSFV, APES$_\text{HARM}$, GTF$_\text{F0}$ and ISE$_\text{ASD}$.}
\label{estoi_boxplots}
\end{figure*}

\section{Experiments and Discussion}

This Section presents a the experiments results and discussion for the proposed ISE$_\text{ASD}$ and competitive methods in terms of intelligibility and quality improvement. Initially, the competitive techniques SSFV, APES$_\text{HARM}$ and GTF$_\text{F0}$ are briefly described. Following, the ESTOI metric is here adopted to examine the predicted intelligibility while PESQ \cite{PESQ_2001} is used to evaluate the quality rates of noisy speech signals. Finally, a listening test \cite{GHIMIRE_2012} is conducted to verify the perceptual differences between NT and ASD groups under noisy conditions. 

\subsection{Competitive Methods}

The baseline techniques adopted in the results comparison are described below:
\vspace{0.3cm}

\subsubsection{\bf SSFV}

The main idea of this solution consists of transforming the original signal adopting a Lombard effect strategy \cite{LOMBARD_1911} \cite{JUNQUA_1993}. In this effect the central frequencies of the formants are shifted (Formant Shifting). It moves away the energy from these frequencies from the region of spectral action of the noise. The formant shifting process is described in \cite{GAEL_2016} and optimized to operate in environments with the presence of Car noise (composed by radio, message alert and telephone). Initially, LPC (Linear Prediction Coding) is used to estimate the poles and formant frequencies of the voiced speech signal. In the LPC model, a 25ms frame of the signal $s(n,m)$ can be represented by linear predictions of order $p$ \cite{RABINER_1978}, that is

 \begin{equation}
  s(n,m)=\sum_{j=1}^{p}a_js(n-j,m)+e(n,m),
 \end{equation}
where $a_j$ are the linear prediction coefficients, $e(n,m)$ indicates the residual error and $p=12$. The variables $n$ and $m$ represent the signal sample and time frame indices, respectively. The LP filter A($z$) is obtained from the coefficients $a_j$, so that $A(z)=1+\sum_{j=1}^{p}a_jz^j$. The poles \textbf{P} are obtained by the roots of the LP coefficients, and the formant frequencies \textbf{F} are defined as the estimated pole angles.

The formants obtained are shifted according to a function $\delta(F)$ \cite{GAEL_2017} related to the characteristics of the acoustic noise. The displacement of formants is carried out according to the criterion

 \begin{equation}
    \hat{F}(f) = \left\{\begin{matrix}
    F(f)+\delta(f), \quad f_1<f<f_3 \\
    F(f), \quad \quad \quad \quad \text{otherwise}.
    \end{matrix}\right.
 \end{equation}
where $f_1$ and $f_3$ are the first and third formants, respectively. Finally, the resulting set of formants $\hat{\textbf{F}}$ is obtained from these modifications.
\vspace{0.3cm}

\subsubsection{\bf APES$_\text{HARM}$}

This technique \cite{NORHOLM_2016} adopts the APES filter \cite{STOICA_1999} with an harmonic chirp model (APES$_\text{HARM}$). The F0 is estimated in each voiced segment, and the harmonics are defined as its integer multiples ($f_c = F0, 2 F0, \ldots, L F0$). Next, the APES filter bank is implemented, with central frequencies defined by $f_c$. The signal reconstruction is performed using the filtered voiced segments and the original unvoiced (\emph{unvoiced}) segments of the speech signal. The processed signal shows an increase in SNR when compared to the original signal, as the APES filter attenuates the effect of certain spectral regions of the noise.
\vspace{0.3cm}

\subsubsection{\bf GTF$_\text{F0}$}

In the GTF$_\text{F0}$ \cite{QUEIROZ_2021} method, a set of $L$ Gammatone filters $\left\{ h_k(t), k=1 \ldots, L \right\}$ are applied to successively filter the input sample sequence $x_q(t)$. Each filter $h_k(t)$ is implemented in frames of 32 ms considering order $n = 4$, and center frequency

\begin{equation}
 f_c = k F0
\end{equation}
and bandwidth $b = 0.25 F0$. The time-domain impulse response function described in (\ref{eq:gamma_c}) is applied for GTF$_\text{F0}$.

After the Gammatone filtering, the amplitudes of the output samples $y_q^k(t), k = 1, \ldots, L$ are amplified by the following a gain factor $G_k \geq 1$. The integer multiples of F0 are amplified as in \cite{QUEIROZ_2021} with the following linear gains: $G_1$ = $G_2$ = 5.0, $G_3$ = 4.0 and $G_4$ = 2.5.

\subsection{Objective Evaluation: ESTOI}
\vspace{0.1cm}

\begin{table}[t!] \caption{\label{estoiUNP} Average ESTOI values [\%] for Unprocessed Noisy Speech Signals.}
\renewcommand{\arraystretch}{1.7}
\begin{center}
{
\begin{tabular}{lccccc}

\hline
&\multicolumn{4}{c}{SNR (dB)}&\\\cline{2-5}

Noise&-10&-5&0&5&Average\\\hline

Babble (INS$_\text{max}$=34.6)&17.5&28.1&40.3&53.4&34.8\\
Cafeteria (INS$_\text{max}$=11.7)&19.4&30.2&42.8&56.6&37.2\\
Helicopter (INS$_\text{max}$=1.9)&29.1&40.0&51.7&64.1&46.2\\
SSN (INS$_\text{max}$=1.2)&17.2&27.9&39.8&52.5&34.3\\\hline
Average&20.8&31.6&43.7&56.7&38.1\\\hline

\end{tabular}
}
\end{center}
\end{table}

Several evaluation experiments are conducted considering the test subset attained from \cite{GONZALEZ_2014}, which also provides the V/UV information. This set is composed of 192 speech utterances from TIMIT database \cite{TIMIT_1993} sampled at 16 kHz, spoken by 24 speakers (16 male and 8 female). Each speech segment has on average 3 seconds duration. The training subset adopted for the gain characterization is independent of the test subset in terms of speakers and speech content. Four additive background noises (2 stationary and
2 non-stationary)\footnote{The INS values are available at lasp.ime.eb.br/downloads/LASP\_ISE\_ASD.zip} are selected: Babble from RSG-10 \cite{RSG_1988} database, Cafeteria and Helicopter from Freesound.org\footnote{Available at www.freesound.org.} and SSN \cite{DEMAND_2013}. Four SNR values are considered varying from -10 dB up to 5 dB. In summary, considering 192 signals, 4 noises, 4 SNR values and 4 comparative methods, it is conducted a total of 12,288 experiments.

Table \ref{estoiUNP} presents the average ESTOI results for the unprocessed (UNP) noisy speech signals. The Index of Nonstationarity (INS) \cite{BORGNAT_2018} is computed for each noise source. The INS$_\text{max}$ values indicate that Babble and Cafeteria are nonstationary, while Helicopter and SSN are stationary. Here the SSN and Babble noises attained the lowest scores for SNR value of -10 dB, with of 17.2\% and 17.5\%, respectively. Thus, these conditions denote the most challenging scenarios, and the intelligibility can be considered bad according to the Speech Transmission Index (STI) \cite{STI_1997}\cite{VARY_2006}. On the other hand, the STI scale indicates a fair intelligibility for SNR = 5 dB, except in Helicopter noise (indicated as good).

Fig. \ref{estoi_boxplots} depicts the ESTOI intelligibility improvement results ($\Delta$ESTOI) achieved with the proposed and baseline solutions. Note that the ISE$_\text{ASD}$ attained the highest average improvement in all of the 16 cases. For instance, for the Babble noise with SNR = -10 dB the ISE$_\text{ASD}$ accomplished 8.8 p.p. on ESTOI score, compared to 7.1 p.p., --0.5 p.p. and --2.8 p.p. of GTF$_\text{F0}$, SSFV and APES$_\text{HARM}$ approaches, respectively. Moreover, for the Helicopter scenario with SNR = -5 dB UNP signals have the ESTOI incremented by the proposed method from 0.40 to 0.54. That is, the poor values became a regular intelligibility according to STI.

\begin{figure*}[t!]
\centering
\includegraphics[width=0.55\linewidth,clip=false,trim=0pt 0pt 0pt 0pt]{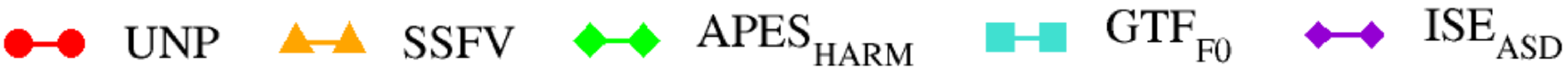}\\\vspace{0.15cm}
\hspace{-0.05cm}\includegraphics[width=0.275\linewidth,clip=true,trim=0pt 0pt 0pt 0pt]{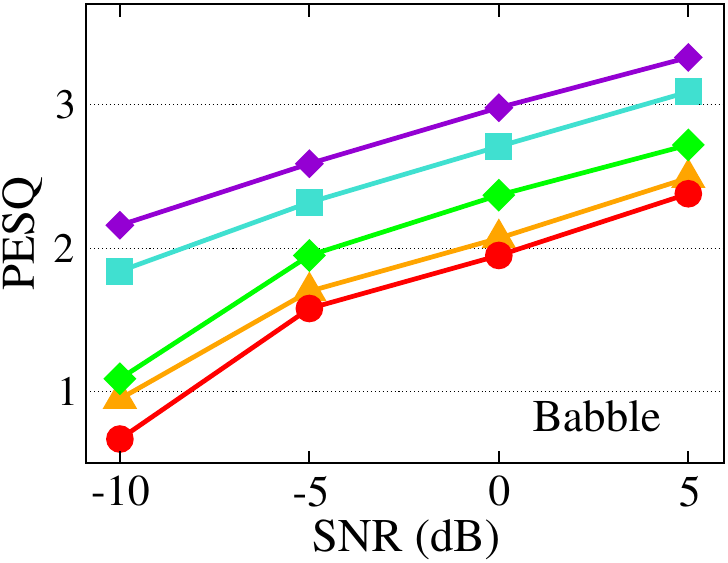}\hspace{-0.12cm}
\includegraphics[width=0.243\linewidth,clip=true,trim=40pt 0pt 0pt 0pt]{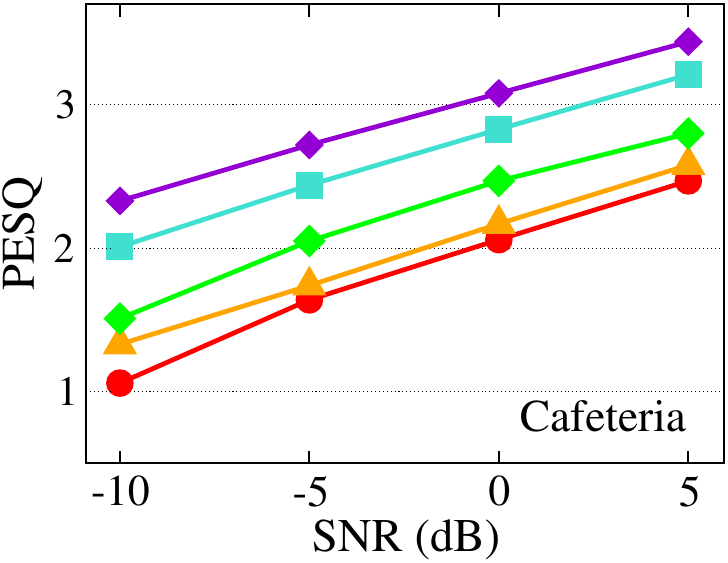}\hspace{-0.11cm}
\includegraphics[width=0.243\linewidth,clip=true,trim=40pt 0pt 0pt 0pt]{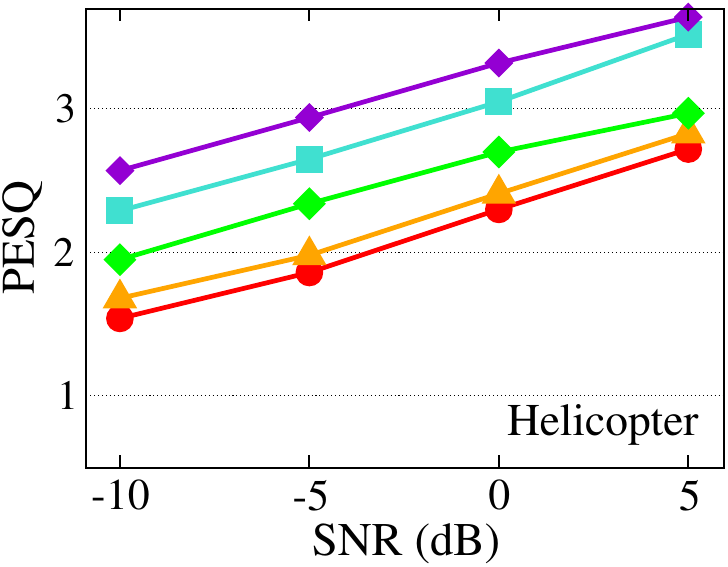}\hspace{-0.12cm}
\includegraphics[width=0.243\linewidth,clip=true,trim=40pt 0pt 0pt 0pt]{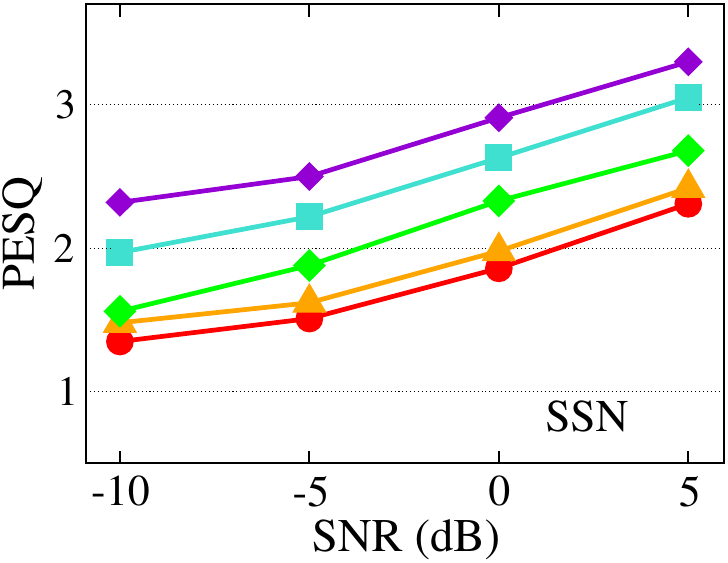}
\caption{Average PESQ results for the noisy speech signals in different SNR values for UNP signals and processed with the competitive methods: SSFV, APES$_\text{HARM}$, GTF$_\text{F0}$ and ISE$_\text{ASD}$.}
\label{pesq_res}
\end{figure*}

An increase of the STI level can also be noted for noisy signals corrupted by Helicopter with 0 dB. In this condition, the average ESTOI is 51.7 for UNP. This value is incremented to 64.2 when processed by the proposed solution. It denotes an update on STI from regular to a good intelligibility. Furthermore, for SNR = 5dB, the ISE$_\text{ASD}$ attains ESTOI scores $>$ 0.75, which is related to be a excellent intelligibility. The APES$_\text{HARM}$ is outperformed by the other approaches in all cases. Finally, the proposal also reaches the highest ESTOI scores for the SSN noise, outperforming in 2.1 p.p. the competitive GTF$_\text{F0}$. Since this noise is used in subjective tests, this improvement also might reflect in the perceptual experiments.

% \begin{table*}[t!]
% \begin{center}
% \caption{PESQ Objective Scores for Noisy Conditions at Different SNR Values}
% {
% \renewcommand{\arraystretch}{1.3}
% \setlength{\tabcolsep}{3.3pt}
% \begin{tabular}{lccccccccccccccccccccccccc}
% \hline
% \multirow{2}{*}&\multicolumn{5}{c}{Babble}&&\multicolumn{5}{c}{Cafeteria}&&\multicolumn{5}{c}{Helicopter}&&\multicolumn{5}{c}{SSN}&&\\
% \cline{2-6}\cline{8-12}\cline{14-18}\cline{20-24}
%
% SNR(dB)&-10&-5&0&5&Av.&&-10&-5&0&5&Av.&&-10&-5&0&5&Av.&&-10&-5&0&5&Av.&&Overall\\\hline
%
% UNP&0.67&1.58&1.95&2.38&1.65&&1.06&1.64&2.06&2.47&1.72&&1.54&1.86&2.30&2.72&2.10&&1.35&1.51&1.86&2.31&1.76&&1.81 \\
%
% SSFV&1.11&1.60&1.97&2.39&1.77&&1.33&1.69&2.07&2.48&1.89&&1.58&1.88&2.31&2.73&2.12&&1.48&1.52&1.88&2.32&1.80&&1.90 \\
%
% APES$_\text{HARM}$&1.09&1.95&2.37&2.72&2.03&&1.51&2.05&2.47&2.80&2.21&&1.95&2.34&2.70&2.97&2.49&&1.56&1.88&2.33&2.68&2.11&&2.21\\
%
% GTF$_\text{F0}$&1.84&2.32&2.71&3.09&2.49&&2.01&2.44&2.83&3.21&2.62&&2.29&2.65&3.05&3.52&2.88&&1.97&2.22&2.63&3.05&2.47&&2.61\\
%
% pGTF$_\text{F0}$&\textbf{2.16}&\textbf{2.59}&\textbf{2.98}&\textbf{3.33}&\textbf{2.77}&&\textbf{2.33}&\textbf{2.72}&\textbf{3.08}&\textbf{3.44}&\textbf{2.89}&&\textbf{2.57}&\textbf{2.94}&\textbf{3.32}&\textbf{3.64}&\textbf{3.12}&&\textbf{2.32}&\textbf{2.50}&\textbf{2.91}&\textbf{3.30}&\textbf{2.76}&&\textbf{2.88}\\\hline
%
% \end{tabular}\label{pesq_res}
% }
% \end{center}
%
% \end{table*}

\subsection{Objective Evaluation: PESQ}

The predicted quality rates computed with the Perceptual Evaluation of Speech Quality (PESQ) \cite{PESQ_2001} are shown in Fig. \ref{pesq_res}. The PESQ score is here computed from 30\% of the most relevant voiced frames of noisy speech. These frames are selected from those with the lowest signal-to-noise ratio values. The results are presented according to the Mean Opinion Score (MOS) scale \cite{MOS_2017} from 1 to 5, with 5 being the best possible score. Note that ISE$_\text{ASD}$ outperforms the competing approaches for most of the noisy speech conditions in terms of quality assessment. For the Babble scenario, this method presents average PESQ of 2.77, followed by 2.49, 2.03 and 1.77 for GTF$_\text{F0}$, APES$_\text{HARM}$, and SSFV baselines, respectively.
\vspace{0.1cm}

It is important to highlight that only ISE$_\text{ASD}$ accomplished more than 1 point in the MOS scale in the overall average results among the four competitive methods. These results reinforce the robustness of the personalized solution in terms of noise effects attenuation. Consequently, it will be reflected in improvements in both acoustic intelligibility and quality, resulting in better acoustic perception mainly in ASD individuals.

\begin{table}[t!] \caption{\label{pValue} Probability p value attained from the Analysis of Variance (ANOVA).}
\renewcommand{\arraystretch}{1.7}
\setlength{\tabcolsep}{3.5pt}
\begin{center}
{
\begin{tabular}{clccccccccc}

\hline

&SNR (dB)&-10&-5&0&5\\\hline

\multirow{4}{*}{ESTOI}&Babble&2.4$e^{-121}$&7.1$e^{-132}$&2.4$e^{-139}$&3.3$e^{-142}$\\
&Cafeteria&7.2$e^{-150}$&3.4$e^{-156}$&4.4$e^{-155}$&4.1$e^{-144}$\\
&Helicopter&3.5$e^{-205}$&6.4$e^{-193}$&8.5$e^{-172}$&4.8$e^{-148}$\\
&SSN&3.2$e^{-177}$&1.2$e^{-189}$&4.9$e^{-187}$&2.2$e^{-165}$\\\hline

\multirow{4}{*}{PESQ}&Babble&1.1$e^{-77}$&3.1$e^{-140}$&1.0$e^{-187}$&1.0$e^{-208}$\\
&Cafeteria&7.7$e^{-101}$&1.6$e^{-159}$&2.1$e^{-214}$&4.4$e^{-240}$\\
&Helicopter&2.7$e^{-139}$&1.3$e^{-188}$&7.9$e^{-234}$&4.0$e^{-265}$\\
&SSN&2.5$e^{-67}$&1.7$e^{-139}$&3.9$e^{-191}$&8.5$e^{-220}$\\\hline

\end{tabular}
}
\end{center}
\end{table}

\subsection{Statistical Significance}

The experiments results are also evaluated according to their statistical significance, considering a significance level of 5\% ($5.0 e^{-2}$). Table \ref{pValue} exibits the probability $p$ values attained with the Analysis of Variance (ANOVA) in all test conditions. The highest $p$ value is achieved by Babble ($p=2.4 e^{-121}$) for ESTOI and SSN noise ($p=2.5 e^{-67}$) for PESQ metric, both with SNR = -10dB. Note that all results present $p \leq 5e^{-2}$, rejecting the null hypothesis of equivalent distributions. Thus, the intelligibility and quality values are statistically significant in all cases.
\vspace{0.1cm}

\subsection{Perceptual Intelligibility Tests}
\vspace{0.3cm}

A subjective intelligibility evaluation is conducted in order to examine the proposed and baseline methods in noisy speech signals. The Telecommunication Standardization Sector (ITU-T) P.863 recomendation is adopted in the perceptual tests of word recognition \cite{GHIMIRE_2012}. The database is composed of phonetically balanced words recorded at the Laboratory of Acoustic Signal Processing (LASP)\footnote{The complete test database is available at lasp.ime.eb.br.}. The recording is conducted with a Shure Beta microphone and a M-Audio fasttrack guitar-mic interface.
\vspace{0.2cm}

Three methods for intelligibility improvement are examined in the first stage of the perceptual tests: SSFV \cite{GAEL_2017}, APES$_\text{HARM}$ \cite{NORHOLM_2016} and GTF$_\text{F0}$ \cite{QUEIROZ_2021}. The test is conducted with 7 male and 7 female Brazilian volunteers for each NT and ASD group with respective ages between [19-57] and [16-34], i.e., a total of 28 volunteers. The SSN noise is adopted to corrupt the words with SNRs of -5 dB, 0 dB and 5 dB. Ten words are applied for each of the 12 test conditions, i.e., three SNR levels and three methods plus the unprocessed (UNP) case. Participants are introduced to the task in a training session with 4 words, in order to adjust the volume. The material is diotically presented using a pair of Roland RH-200S headphones. Listeners hear each word once in Brazilian Portuguese, presented in an arbitrary order and are asked to indicate the word in a sheet list with five options. The average time taken by each volunteer to hear the 150 words of the test and mark the options is 25 minutes. It is important to mention that the time spent for tests of the group with ASD is higher than NT individuals, due to their inherent hypersensitivity.
\vspace{0.2cm}

The intelligibility results for each method are presented in Table \ref{percep_tests}. Observe that female volunteers presented higher intelligibility rates than male with average score of 8.4 p.p. (percentage points) higher for NT listeners. The overall average score of UNP and GTF$_\text{F0}$ presented with NT was 67.6\% and 83.0\%. Note that in the same cases ASD presented lower scores (53.9\% and 73.6\%). Observe that this difference between groups increases for lower SNR values. This indicates a lower acoustic perception of ASD group in this test with SSN noise. Since this evident contrast between NT and ASD, it can be concluded that the subjective listening test might serve as an auxiliary tool to detect ASD behavior.
\vspace{0.2cm}

\begin{table}[t!] \caption{\label{percep_tests} Perceptual Tests Average Results [\%] for NT and ASD Volunteers.}
\renewcommand{\arraystretch}{1.5}
\setlength{\tabcolsep}{3.5pt}
\begin{center}
{
\begin{tabular}{clccccccccc}

\hline
&\multirow{2}{*}&\multicolumn{4}{c}{NT}&&\multicolumn{4}{c}{ASD}\\
\cline{3-6}\cline{8-11}

&SNR (dB)&-5&0&5&Av.&&-5&0&5&Av.\\\hline

\multirow{5}{*}{Male}&UNP&52.0&67.5&81.8&67.1&&27.3&55.8&71.4&51.5\\
&SSFV&26.0&57.1&77.9&53.7&&16.9&46.8&74.0&45.9\\
&APES$_\text{HARM}$&18.2&29.9&61.0&36.4&&10.4&23.4&45.5&26.4\\
&GTF$_\text{F0}$&68.8&79.2&93.5&80.5&&55.8&71.4&94.8&74.0\\\hline

\multirow{5}{*}{Female}&UNP&54.6&68.2&83.8&68.9&&51.5&54.6&72.7&59.6\\
&SSFV&24.2&66.7&90.9&60.6&&15.2&48.5&66.7&43.4\\
&APES$_\text{HARM}$&15.2&21.2&66.7&34.3&&12.1&36.4&48.5&32.3\\
&GTF$_\text{F0}$&81.8&84.9&100&88.9&&57.6&72.7&87.9&72.7\\\hline

\multirow{5}{*}{Overall}&UNP&52.7&67.7&82.4&67.6&&34.6&55.5&71.8&53.9\\
&SSFV&25.5&60.0&81.8&55.8&&16.4&47.3&71.8&45.2\\
&APES$_\text{HARM}$&17.3&27.3&62.7&35.8&&10.9&27.3&46.4&28.2\\
&GTF$_\text{F0}$&72.7&80.9&95.5&83.0&&56.4&71.8&92.7&73.6\\\hline

\end{tabular}
}
\end{center}
\end{table}

\begin{table}[t!] \caption{\label{percep_ise} Perceptual Tests Results [\%] for ISE$_\text{ASD}$.}
\renewcommand{\arraystretch}{1.5}
\setlength{\tabcolsep}{3.5pt}
\begin{center}
{
\begin{tabular}{lccccccccc}
\hline

&\multicolumn{4}{c}{NT}&&\multicolumn{4}{c}{ASD}\\
\cline{2-5}\cline{7-10}
SNR (dB)&-5&0&5&Av.&&-5&0&5&Av.\\\hline
Male&77.1&87.1&94.3&86.2&&68.8&74.0&93.5&78.8\\
Female&84.3&88.6&98.6&90.5&&63.6&85.7&96.1&81.8\\
Overall&80.7&87.9&96.5&88.4&&66.2&79.9&94.8&80.3\\\hline

\end{tabular}
}
\end{center}
\end{table}

Futher perceptual tests were conducted to evaluate the intelligibility attained by the proposed solution. The same amount of volunteers and words are considered in this additional experiments. Table \ref{percep_ise} presents the results for ISE$_\text{ASD}$ for the NT and ASD volunteers. For most cases, ISE$_\text{ASD}$ outperforms the competitive approaches. Note that the proposed solution attained overall average values 5.5 p.p. and 7.6 p.p. higher than the competitive GTF$_\text{F0}$ for NT and ASD volunteers (see Table \ref{percep_tests}), respectively. In comparison to the objective results where the ESTOI gain of ISE$_\text{ASD}$ is 2.1 p.p higher (see Fig. \ref{estoi_boxplots}) than this baseline. Hence, it can be observed that the subjective results evidenced the intelligibility accomplished by the proposed scheme.
\vspace{0.2cm}

\subsection{Normalized Processing Time}
\vspace{0.3cm}

Table \ref{processing_time} indicates the computational complexity which refers to the normalized processing time required for each method evaluated for 512 samples per frame, i.e., 32 ms). These values are obtained with an Intel (R) Core (TM) i7-9700 CPU, 8 GB RAM, and are normalized by the execution time of the proposed ISE$_\text{ASD}$ solution. The processing time required for fundamental frequency estimation is also considered here. Note that the APES$_\text{HARM}$ scheme presents the longest processing time, followed by ISE$_\text{ASD}$ and GTF$_\text{F0}$ since the HHT-Amp estimation are based on the EEMD, and demand a relevant computational cost.
\vspace{0.3cm}

\begin{table}[t!] \caption{\label{processing_time} Normalized Mean Processing Time.}
\renewcommand{\arraystretch}{1.5}
\begin{center}
{
\begin{tabular}{cccc}

\hline
SSFV&APES$_\text{HARM}$&GTF$_\text{F0}$&ISE$_\text{ASD}$\\\hline

0.32&1.14&0.93&1.00\\ \hline
\end{tabular}
}
\end{center}
\end{table}

\section{Conclusion}
\vspace{0.35cm}

This paper introduced a novel method to improve intelligibility for ASD situation under urban noisy environment. It is composed of three main steps. First, the HHT-Amp technique is adopted to estimate the F0 from voiced frames. Then, a Gammatone filterbank was applied to the frames considering multiple harmonic frequency bands according to the F0. Finally, the filtered components were amplified by gain factors in order to emphasize the harmonic features of speech. Extensive experiments were conducted to evaluate the intelligibility and quality enhancement provided by the ISE$_\text{ASD}$ method and competitive approaches. Four acoustic noises were considered to compose the noisy scenarios. Two objective measures were used to evaluate the competitive solutions. Results showed that ISE$_\text{ASD}$ outperforms the baseline methods in terms of intelligibility and quality scores for all the acoustic noises. Perceptual tests proved to be a useful tool to assist in detection of ASD. Besides that, those results indicate that the proposed solution outperforms the baselines in terms of subjective intelligibility imprevement. Further research includes the exploitation of the identified impairment due to the composition of HIN and urban acoustic noise solution to serve as an additional aid to ASD diagnosis.
\vspace{0.2cm}

% Can use something like this to pu references on a page
\ifCLASSOPTIONcaptionsoff
  \newpage
\fi



%\bibliographystyle{IEEEtran}


%   0.5em minus 0.4em\relax Harlow, England: Addison-Wesley, 1999

% \end{thebibliography}

\bibliographystyle{ieeetr}
\bibliography{tasl}

\end{document}